\documentclass{article}
\usepackage[utf8]{inputenc}
\usepackage{graphicx}
\usepackage{amsmath}
\usepackage[version=4]{mhchem}
\usepackage{siunitx}
\usepackage{tabularx}
\usepackage{longtable}
\usepackage{float}
\usepackage{placeins} 
\usepackage{xcolor}
\usepackage{subfigure}
\usepackage{amsmath}
\usepackage[sort&compress,numbers]{natbib}
\usepackage{hyperref}
\usepackage[letterpaper,margin=1in]{geometry}

\begin{document}

\title{Aeroelastic Analysis of Transonic Flutter with CFD-Based Reduced-Order Model}
\author{Ana Cristina Neves Carloni\footnote{Instituto Tecnológico de Aeronáutica, 12228-900 São José dos Campos, SP, Brazil} \and João Luiz F. Azevedo\footnote{Instituto de Aeronáutica e Espaço, 12228-904 São José dos Campos, SP, Brazil}}
\date{}
\vspace{-.5cm}
\maketitle

\begin{abstract}
The current work is concerned with studying processes for constructing reduced-order models capable of performing transonic aeroelastic stability analyses in the frequency domain based on computational fluid dynamics (CFD) techniques. The CFD calculations are based on the Euler equations, and the code uses a finite volume formulation for general unstructured grids. A centered spatial discretization with added artificial dissipation is used, and an explicit Runge-Kutta time marching method is employed. The dynamic system considered in the present work is a NACA 0012 airfoil-based typical section in the transonic regime. Unsteady calculations are performed for mode-by-mode and simultaneous excitation approaches, the latter defined by orthogonal Walsh functions. System identification techniques are employed to allow the splitting of the aerodynamic coefficient time histories into the contribution of each individual mode to the corresponding aerodynamic transfer functions. Generalized unsteady aerodynamic forces are approximated by a rational transfer function in the Laplace domain, in which nonlinear parameters are selected through a non-gradient optimization process. Different types of interpolating polynomials are tested and the results are compared. Results show that the proposed procedure can reproduce literature aeroelastic analysis data with a single unsteady CFD run.
\end{abstract}

\section{Nomenclature}

{\renewcommand\arraystretch{1.0}
\noindent\begin{longtable}{@{}l @{\quad=\quad} l@{}}
$a_{h}$  & distance from mid-chord to the elastic axis normalized by $b$ \\
$b$  & semi-chord length \\
$c$  & airfoil chord length \\
$C_{l}$  & lift coefficient\\
$C_{m}$  & moment coefficient\\
$G_{i,j}$ & transfer function of the input $j$ to the output $i$ \\
$h$  & vertical translation \\
$m$  & airfoil mass \\
$M_{\infty}$ & freestream Mach number \\
$n_{a}$ & number of added aerodynamic state variables \\ 
$r_{\alpha}$  & airfoil dimensionless radius of gyration about the elastic axis \\
$\Bar{s}$ & dimensionless Laplace transform complex variable \\
$t$  & time\\
$U^{*}$  & characteristic speed \\
$U_{\infty}$  & freestream speed\\
$\{x(\Bar{t})\}$ & system state vector \\
$x_{\alpha}$  & distance from the elastic axis to the center of mass normalized by $b$ \\
$\alpha$  & pitch mode coordinate \\
$\alpha_{0}$ & initial angle of attack \\
$\eta$  & generalized coordinate\\
$\kappa$  & reduced frequency\\
$\mu$  & mass ratio \\
$\omega_h$  & uncoupled natural circular frequency of the plunge mode \\
$\omega_\alpha$  & uncoupled natural circular frequency of the pitch mode \\
$\rho_{\infty}$  & freestream density\\
$\xi$  & plunge mode coordinate, where $\xi=h/b$ \\
\end{longtable}}

\section{Introduction}

The field of computational fluid dynamics (CFD) is constantly evolving since its introduction as a branch of numerical methods and applied mathematics. Despite reaching a mature stage after many developments throughout the years, the accurate and affordable simulation of transonic aeroelastic systems is still an open challenge. In the past decades, traditional aeroelastic analyses implement an explicit coupling of aerodynamic and dynamic-structural systems by carrying out an interactive process \cite{silva2009development,camilo2013hopf,silva2004development,silva2008simultaneous,silva2014evaluation}. Accordingly, parametric and flight condition variations necessarily demand a repetitive use of high-fidelity CFD codes. Given the available computational power, the traditional aeroelastic approach is prohibitive for engineering applications that require multiple structural modes. Even nowadays, aircraft designs with large elongation, flexibility, and complexity usually compromise traditional aeroelastic analysis feasibility \cite{waite2019reduced,skujins2014reduced}. As a consequence, this impacts the structural stability condition mapping, which is critical for preventing the occurrence of aeroelastic phenomena, such as flutter \cite{bisplinghoff1955aeroelasticity}. One of the most prominent approaches to tackle this challenge is the reduced-order model (ROM) formulation, where the prevailing dynamics of the aeroelastic system is captured. The present effort builds upon previous work \cite{marques2008numerical,marques2008z,azevedo2012efficient,azevedo2013effects} that take advantage of this approach in transonic aeroelastic applications. In essence, the objective of the present work is to improve the procedure developed to analyze transonic aeroelastic phenomena using reduced-order models in terms of accuracy. Ultimately, the intent is to expand the ROM formulation to handle large deflections that lead to significant changes in the average nonlinear aerodynamic flow. The effects of structural nonlinearities are notably part of the aircraft industry routine due to increasingly flexible commercial aircrafts. This raises the need to develop reduced-order models capable of effectively supporting complex aeroelastic analyses.

The CFD code used here is based on the 2-D Euler equations, which are discretized using a finite volume approach for unstructured grids. A centered scheme with added artificial dissipation is used for spatial discretization, and explicit Runge-Kutta method is employed for time marching. The present paper concentrates initially in identifying the aerodynamic transfer functions from a unique excitation in all the natural modes of the typical section model. This identification procedure is based on the computation of power spectral densities of the inputs and outputs of the dynamic system. A relatively simple approach to estimate the transfer functions is by a division in the frequency domain for uncorrelated input signals and its derivatives. The effects of using different signal processing configurations to identify the transfer functions are studied. Once the transfer functions are obtained, they can be estimated by a rational-function approximation (RFA) in the Laplace domain \cite{tiffany1987nonlinear,eversman1991consistent} in order to better suit the solution of the aeroelastic eigenvalue problem. Finally, the solution of such eigenvalue problem for varying dynamic pressures yields a root locus from which one can estimate the flutter boundary of the configuration.

\section{Theoretical Formulation}

\subsection{General Formulation}

\begin{figure}[hbt!]
\centering
\includegraphics[width=.42\textwidth]{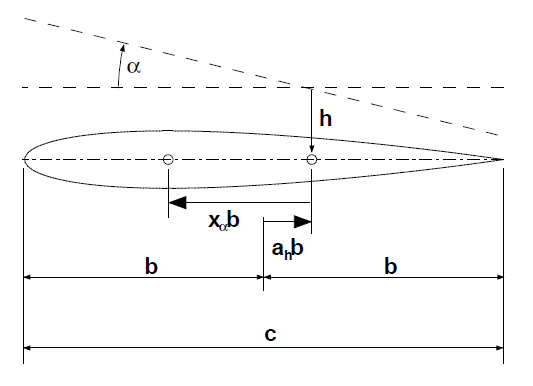}
\caption{\label{fig:figure1}Typical section configuration.}
\end{figure}

The structural model considered in the present work is the typical section, which is widely known and reported in the literature \cite{bisplinghoff1955aeroelasticity,marques2008z,azevedo2013effects}. The dynamic system represented in the typical section is a rigid airfoil with two degrees of freedom, plunge and pitch, as shown in Fig.\ \ref{fig:figure1}. As presented in Fig.\ \ref{fig:figure1}, $h$ is the vertical translation, positive downwards, and $\alpha$ is the pitch mode coordinate, positive in the nose-up direction. The mesh used in the present work is the same as in Ref.\ \cite{azevedo2013effects}, which has about 16,000 triangular control volumes and just over 290 points along the airfoil. It provides adequate resolution of the phenomena of interest when considering an inviscid formulation. The considered mesh movement takes into account the body motions involved in unsteady calculations by rigidly moving the mesh accordingly. In this approach, the far-field boundary conditions are adjusted to account for the boundary movement with the rest of the mesh.

A general aeroelastic system is characterized by aerodynamic, elastic and inertial forces dynamically interacting with structural deformations \cite{bisplinghoff1955aeroelasticity}. It is very common to represent the aerodynamic effects exclusively through the resulting forces and moments acting on the structure as a forcing term. As such, the governing equations for this dynamic system are given by
\begin{equation}
    \label{eq:equation1}
    [M] \{\Ddot{\eta}(t)\} + [K]\{\eta(t)\} = \{Q_{a}(t)\}
\end{equation}
\noindent where the generalized mass and stiffness matrices are, respectively,
\begin{equation}
    [M] = \begin{bmatrix}
            1 & x_{\alpha} \\
            x_{\alpha} & r_{\alpha}^2 
          \end{bmatrix}
\quad \text{and} \quad
    [K] = \begin{bmatrix}
            \omega_{h}^2 & 0 \\
            0 & r_{\alpha}^2 \omega_{\alpha}^2 
           \end{bmatrix} ,
\end{equation}
\noindent and the generalized coordinate and generalized aerodynamic forces are
\begin{equation}
    \{\eta(t)\} = \begin{Bmatrix}
                    \xi(t)  \\
                    \alpha(t) 
                  \end{Bmatrix}
\quad \text{and} \quad
    \{Q_{a}(t)\} = \begin{Bmatrix}
            \dfrac{Q_{a_h}(t)}{mb} \\
            \dfrac{Q_{a_\alpha}(t)}{mb^2}
          \end{Bmatrix} .
\end{equation}

It is convenient to nondimensionalize the aeroelastic equation in order to represent more general situations. This is accomplished with the procedure proposed in Ref.\ \cite{oliveira1993metodologia}, using a reference circular frequency to nondimensionalize the time variable,
\begin{equation}
    \Bar{t} = \omega_{r} t.
\end{equation}
\noindent With the application of the chain rule for the time derivatives, Eq.\ \ref{eq:equation1} becomes
\begin{equation}
    [M] \{\Ddot{\eta}(\Bar{t})\} + [\Bar{K}]\{\eta(\Bar{t})\} = \{\Bar{Q}_{a}(\Bar{t})\}
\end{equation}
\noindent where
\begin{equation}
    [\Bar{K}] = \dfrac{1}{\omega_{r}^2} [K] 
    \quad \text{and} \quad
    \{\Bar{Q}_{a}(\Bar{t})\} = \dfrac{1}{\omega_{r}^2} \{Q_{a}(t)\} .
\end{equation}

Part of the present objective is to efficiently determine the generalized aerodynamic force vector, $\{\Bar{Q}_{a}(\Bar{t})\}$, for an arbitrary structural behavior. However, due to the nonlinearities of the aerodynamic equations, it is very difficult to obtain a general expression for the aerodynamic response. However, this problem is simplified by extending linearity concepts present in the formulation of the potential aerodynamic equations. As presented in Ref.\ \cite{bisplinghoff1955aeroelasticity}, the linear aerodynamic responses can be individually determined for each mode and, then, superimposed for more general responses. Based on these ideas, Ref.\ \cite{oliveira1993metodologia} proposed the assumption of linearity of the aerodynamic response in the transonic regime with regard to the modal motion. As there are no rigorous linearization procedures involved, there are no guarantees that such assumption holds. But, it is natural to expect this sort of linear hypothesis to be valid at least for very small amplitudes of motion. Actually, as it is shown in Ref.\ \cite{marques2008z}, there is a certain amplitude range in which this hypothesis holds. Furthermore, the aeroelastic phenomena analyzed in the present work are restricted to small amplitude motions.

As a consequence of the linearity assumptions, it is possible to determine the aerodynamic response to a general structural behavior from the convolution of an impulsive or indicial aerodynamic solution \cite{bisplinghoff1955aeroelasticity,silva2004development}. The
convolution operation, however, is more easily handled in the frequency domain, in which it is represented by a simple multiplication operation. Hence, the aerodynamic forces linearized with regard to the modal displacements can be written in the frequency domain. After some mathematical manipulations, it is possible to define the generalized forces as
\begin{equation}
    \{Q_{a} (\kappa)\} = \dfrac{(U^{*})^2}{\pi \mu} [A(\kappa)] \{\eta(\kappa)\}
\end{equation}
\noindent where the mass ratio and the characteristic speed are, respectively,
\begin{equation}
    \mu = \dfrac{m}{\pi \rho_{\infty} b^2} \quad \text{and} \quad  U^{*} = \dfrac{U_{\infty}}{b \omega_r},
\end{equation}
\noindent and the aerodynamic influence coefficient matrix is given by
\begin{equation}
    [A(\kappa)] = \begin{bmatrix}
            -C_{l,h}(\kappa)/2 & -C_{l,\alpha}(\kappa) \\
             C_{m,h}(\kappa) & 2 C_{m,\alpha}(\kappa)
          \end{bmatrix} .
\end{equation}

\subsection{Orthogonal Walsh Functions}

There are different approaches that can be used to determine the aerodynamic coefficients that compose the aerodynamic influence coefficient matrix. As there is no closed-form solution for the compressible subsonic and transonic unsteady aerodynamic forces, this work is aimed at numerical methods in which the impulsive solution is known for a determined number of discrete reduced frequency values. All the aerodynamic responses presented here are obtained by exciting the aerodynamic system on a frequency range of interest through the individual application of a step-like motion in both modes. The applied input signals particularly correspond to the Walsh functions, which are a set of block step functions that form an orthogonal basis of the square-integrable functions. Each function takes positive and negative unitary values, and each step block period is an integer multiple of a $2^n$ division of the function length. Each of these functions may be seen as a line, or a column, of a Hadamard \cite{hadamard93} matrix of order $2^n$. This family of functions is similar to step inputs and, therefore, embodies their impulsive nature with regard to frequency bandwidth. Moreover, one should be careful when choosing the length of the Walsh function so that each block period is capable of exciting the relevant frequencies in the system.

\section{State Space Analysis}

\subsection{General Formulation}

A state space representation of a system corresponds to the description of the system dynamics in terms of first-order differential equations, which may be combined into a first-order vector-matrix differential equation. In the present case, this formulation can be achieved by defining
\begin{equation}
\begin{aligned}
    \{x(\Bar{t})\} &= \begin{Bmatrix} 
                     \{\Dot{\eta}(\Bar{t})\} \\
                      \{\eta(\Bar{t})\}
                       \end{Bmatrix} .
\end{aligned}
\end{equation}
\noindent Hence, the governing equation of motion becomes
\begin{equation}
    \label{eq:equation12}
    [\Tilde{M}] \{\Dot{x}(\Bar{t})\} + [\Tilde{K}]\{x(\Bar{t})\} = \{\Tilde{q}(\Bar{t})\},
\end{equation}
\noindent where
\begin{equation}
    [\Tilde{M}] = \begin{bmatrix}
            [M] & [0_{2 \times 2}] \\
            [0_{2 \times 2}] & [I_{2 \times 2}] 
          \end{bmatrix} ,
\quad
    [\Tilde{K}] = \begin{bmatrix}
            [0_{2 \times 2}] & [\Bar{K}] \\
            -[I_{2 \times 2}] & [0_{2 \times 2}] 
           \end{bmatrix} ,
\quad \text{and} \quad           
   \{\Tilde{q}(\Bar{t})\} = \begin{Bmatrix}
                            \{\Bar{Q}_{a}(\Bar{t})\} \\
                            \{0_{2 \times 1}\}
  \end{Bmatrix} .
\end{equation}

As the generalized aerodynamic forces can be conveniently represented in the frequency domain, the system can be more easily studied in the Laplace domain. Applying the Laplace transform to Eq.\ \ref{eq:equation12}, one obtains
\begin{equation}
    \Bar{s} [\Tilde{M}] \{X(\Bar{s})\} + [\Tilde{K}]\{X(\Bar{s})\} = \{\Tilde{Q}(\Bar{s})\},
\end{equation}
\noindent where
\begin{equation}
    \Bar{s} = \dfrac{s}{\omega_r}
\end{equation}
\noindent is the dimensionless Laplace transform complex variable. The idea behind this procedure is to evaluate the aerodynamic influence coefficient matrix over a reduced frequency range of interest and, by making use of the analytical continuation principle, to extend such result to the entire complex plane.

Regardless of which polynomial is used for the representation of the aerodynamic response in the Laplace domain, the resulting state-space representation of the typical section aeroelastic system has always the form
\begin{equation}
    \{\Dot{\chi}(\Bar{t})\} = [D] \{\chi(\Bar{t})\},
\end{equation}
\noindent where $\{\chi(\Bar{t})\}$ is the new state vector that results from the addition of the aerodynamic state variables. A detailed formulation of the stability matrix of the system, $[D]$, can be found in Refs. \cite{eversman1991consistent,azevedo2013effects,marques2008z}. Finally, the aeroelastic stability analysis can, then, be reduced to the classical eigenvalue problem for each value of the $U^{*}$ parameter,
\begin{equation}
    ([D] - \Bar{s} [I_{N \times N}]) \{\chi(\Bar{s})\} = \{0_{N \times 1}\}
\end{equation}
\noindent where $N= 4+2 n_{a}$ for an aeroelastic system with two structural degrees of freedom and $n_{a}$ aerodynamic states.

\subsection{Eversman and Tewari Polynomials}

The available aerodynamic responses in the frequency domain consist of sets of numerical discrete values, which are not convenient for the solution of Eq.\ \ref{eq:equation12}. One possible solution is to approximate these data using interpolating polynomials. There are a number of different polynomials reported in the literature \cite{tiffany1987nonlinear,eversman1991consistent} from which the interpolating polynomial proposed by Eversman and Tewari \cite{eversman1991consistent} without any provision for the treatment of repeated, or very close, poles is initially selected by the authors for implementation and tests. This choice is based on the fact that such polynomials are the most commonly used in applications similar to the ones intended in the present work, and that they are conveniently constructed for the formulation of aerodynamic state variables. In the Laplace domain, this rational-function approximation is written as
\begin{equation}
    \label{eq:equation16}
    [A(\Bar{s})] = [A_{0}] + [A_{1}] \dfrac{\Bar{s}}{U^{*}} + [A_{2}] \left( \dfrac{\Bar{s}}{U^{*}} \right)^{2} + \sum_{n=1}^{n_{\beta}} \left( [A_{n+2}] \dfrac{U^{*}}{\Bar{s}+U^{*} \beta_{n}}  \right),
\end{equation}
\noindent where the $\beta_{n}$'s are the poles that introduce the aerodynamic lags with respect to the structural modes. The polynomial coefficients, $[A_n]$ and $\beta_n$, are determined by an optimized non-gradient least-squares approximation method known as Nelder-Mead algorithm \cite{nelder1965}. In this case, the number of augmented states is equal to the number of added poles, \textit{i.e.}, $n_{a}=n_{\beta}$. The formulation presented in Eq.\ \ref{eq:equation16} is referred to as the first form of the Eversman and Tewari interpolating polynomials. 

The second form of the Eversman and Tewari polynomials consistently accounts for cases where the optimized values of two or more poles of the approximation are close to one another. Although these poles do not necessarily have to be identical, it is referred to as the phenomenon of repeated poles. Whenever this happens, the linear coefficients, $[A_{n}]$, corresponding to these lag parameters are very close in magnitude, very large, and of opposite signs. Such phenomenon makes the subsequent eigenvalue problem poorly conditioned. To overcome such difficulty, Ref.\ \cite{eversman1991consistent} suggests a slight modification to the lag terms in such situations. When two poles are very close to each other, Ref.\  \cite{eversman1991consistent} shows that these poles can be represented by only one of them. However, this pole must appear in two terms of the polynomial, one linear and one quadratic. The same type of analysis can be extended in the case of three, four, or actually any number of poles that occur very close to each other in the original formulation. Moreover, Ref.\ \cite{eversman1991consistent} demonstrates that this procedure does not alter the fitting accuracy of a given polynomial, but it produces a better-conditioned eigenvalue problem. Therefore, the following approximating polynomial could be a more useful model than the original one given in Eq.\ \ref{eq:equation16}
\begin{equation}
\begin{aligned}
    \relax [A(\Bar{s})] = & [A_{0}] + [A_{1}] \dfrac{\Bar{s}}{U^{*}} + [A_{2}] \left( \dfrac{\Bar{s}}{U^{*}} \right)^{2} + \sum_{n=1}^{n_{\beta}} \left( [A_{n+2}] \dfrac{U^{*}}{\Bar{s}+U^{*} \beta_{n}}  \right) + \sum_{n=n_{\beta_1}+1}^{n_{\beta}} \left( [A_{n+N_{2}+2}] \dfrac{({U^{*}})^{2}}{(\Bar{s}+U^{*} \beta_{n})^{2}}  \right) \\
    + & \sum_{n=n_{\beta_2}+1}^{n_{\beta}} \left( [A_{n+N_{3}+2}] \dfrac{({U^{*}})^{3}}{(\Bar{s}+U^{*} \beta_{n})^{3}}  \right) + \sum_{n=n_{\beta_3}+1}^{n_{\beta}} \left( [A_{n+N_{4}+2}] \dfrac{({U^{*}})^{4}}{(\Bar{s}+U^{*} \beta_{n})^{4}}  \right) + \dots
\end{aligned}
\end{equation}
\noindent with
\begin{equation}
    N_{2}=n_{\beta}-n_{\beta_1}, \quad N_{3}=2 n_{\beta}-n_{\beta_1}-n_{\beta_2}, \quad N_{4}=3n_{\beta}-n_{\beta_1}-n_{\beta_2}-n_{\beta_3}, \dots
\end{equation}
\noindent where it is assumed that the $\beta_1, \dots, \beta_{n_{\beta_1}}$ are non-repeated poles, $\beta_{(n_{\beta_1}+1)}, \dots, \beta_{n_{\beta_2}}$ are poles that occur twice, $\beta_{(n_{\beta_2}+1)}, \dots, \beta_{n_{\beta_3}}$ are poles that occur three times, and so on.

\section{Results and Discussion}

\subsection{Simulation Procedure}

\begin{figure}[b!]
    \begin{center}
    	 \subfigure[First set of Walsh function inputs (WF1).]{ \includegraphics[scale=0.28,trim = 1.5cm 6cm 2cm 7cm,clip]{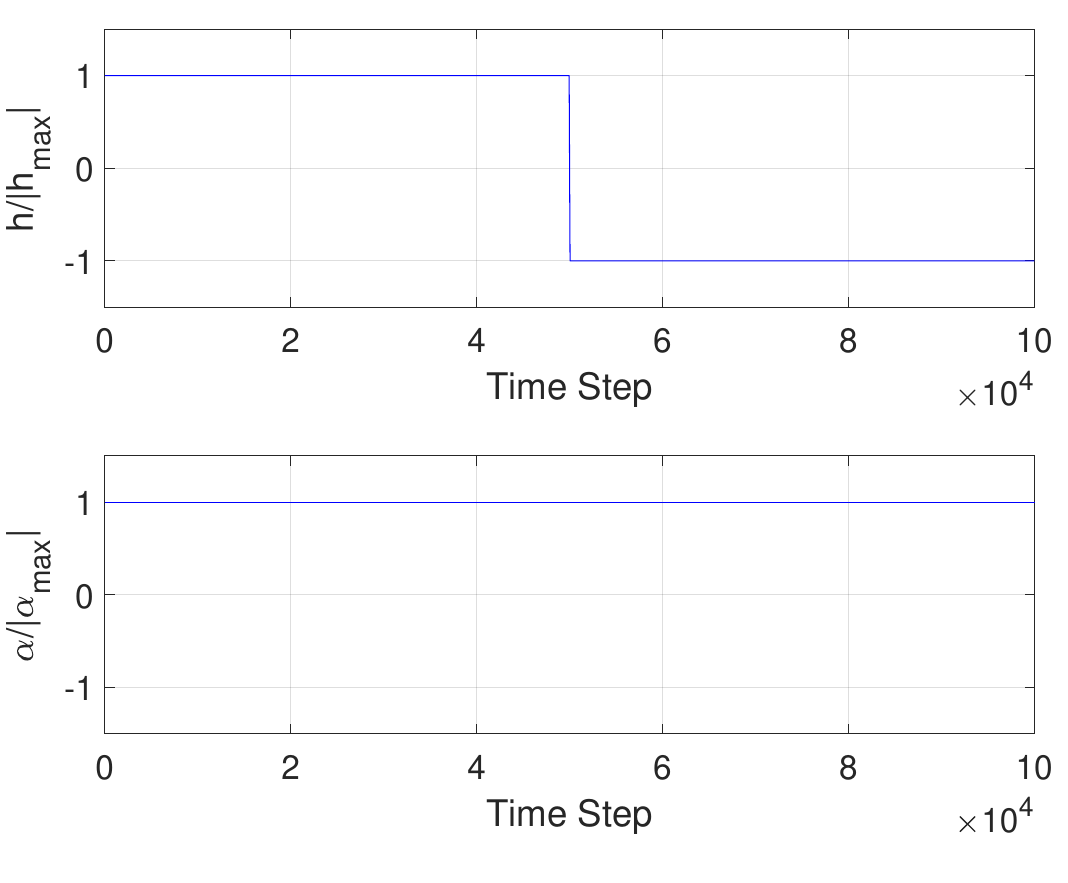}} \quad
    	 \subfigure[Second set of Walsh function inputs (WF2).]{ \includegraphics[scale=0.28,trim = 1.5cm 6cm 2cm 7cm,clip]{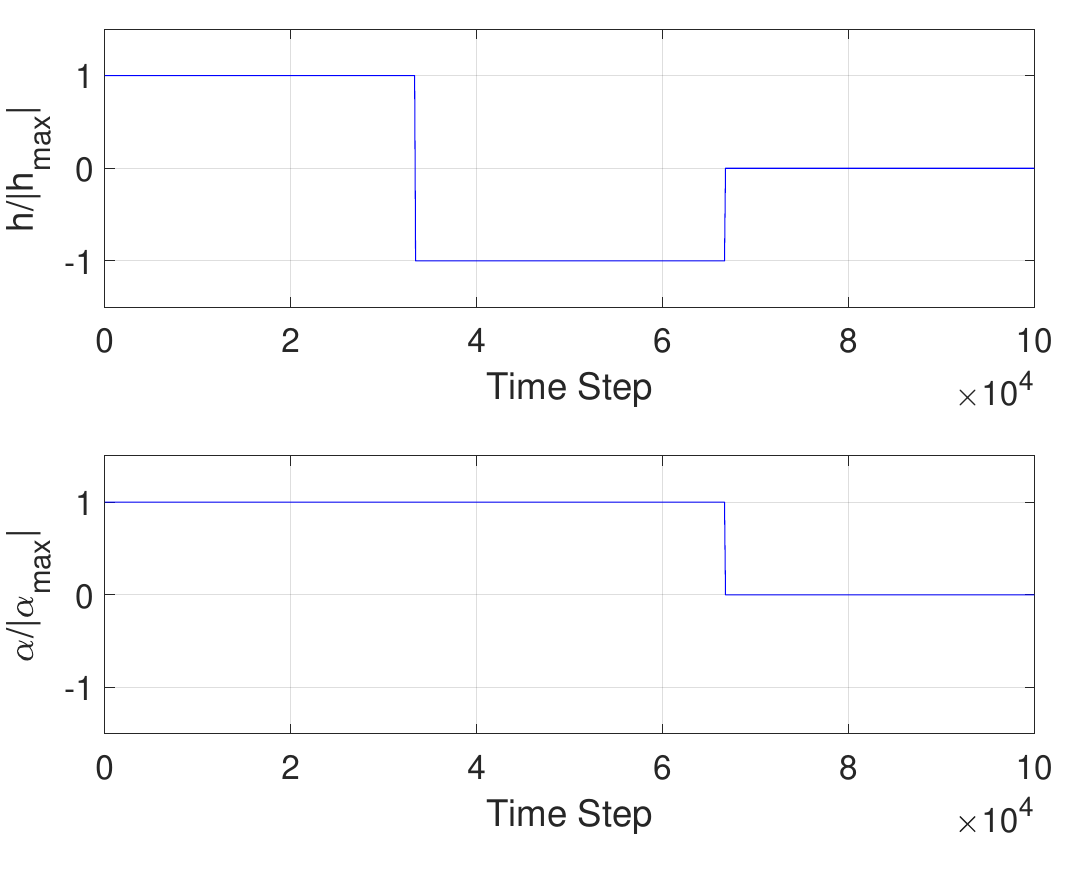}} \quad
    	 \subfigure[Third set of Walsh function inputs (WF3).]{ \includegraphics[scale=0.28,trim = 1.5cm 6cm 2cm 7cm,clip]{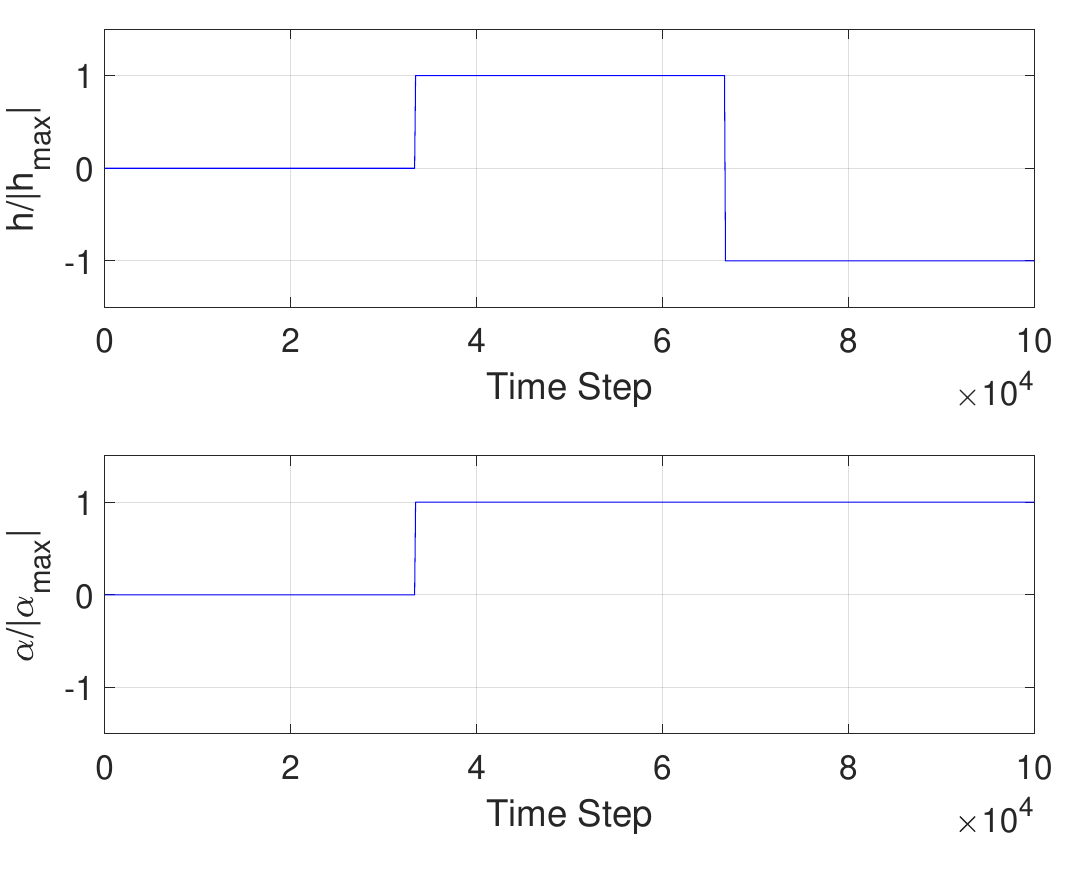}} \quad
    	\subfigure[Fourth set of Walsh function inputs (WF4).]{ \includegraphics[scale=0.28,trim = 1.5cm 6cm 2cm 7cm,clip]{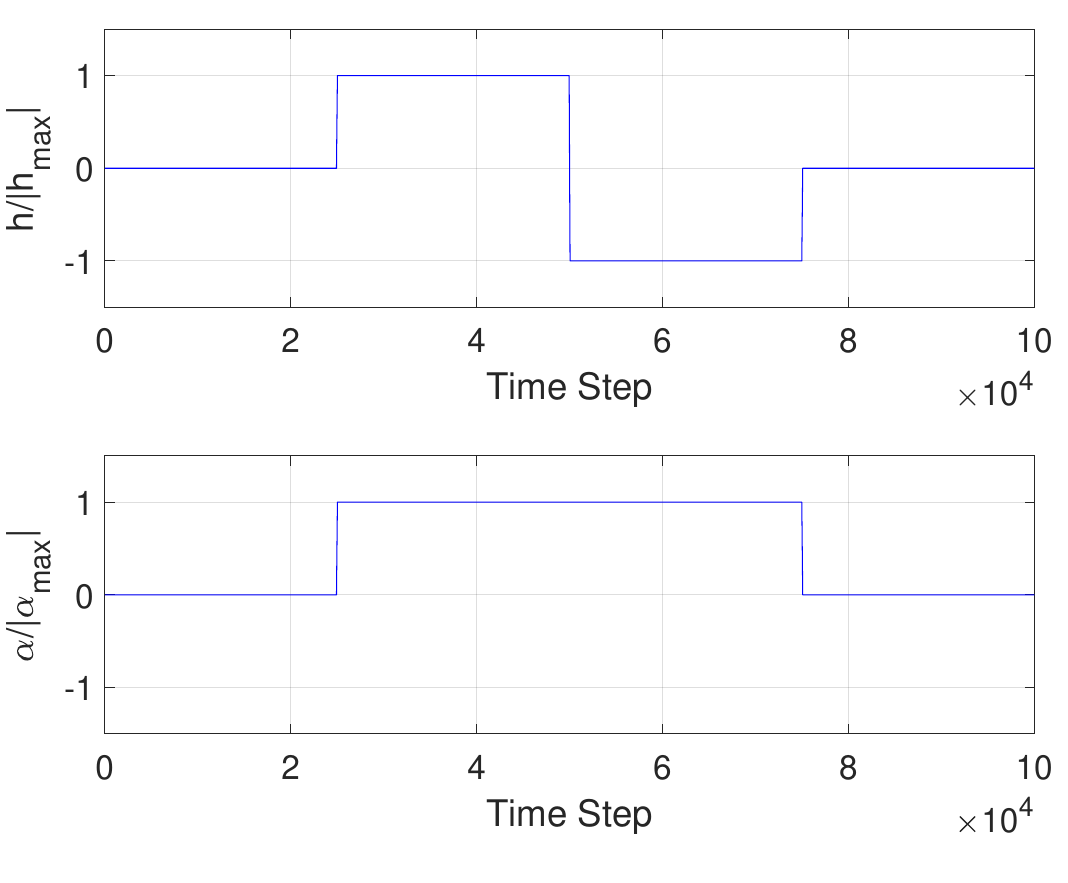}} \quad
    	 \subfigure[Fifth set of Walsh function inputs (WF5).]{ \includegraphics[scale=0.28,trim = 1.5cm 6cm 2cm 7cm,clip]{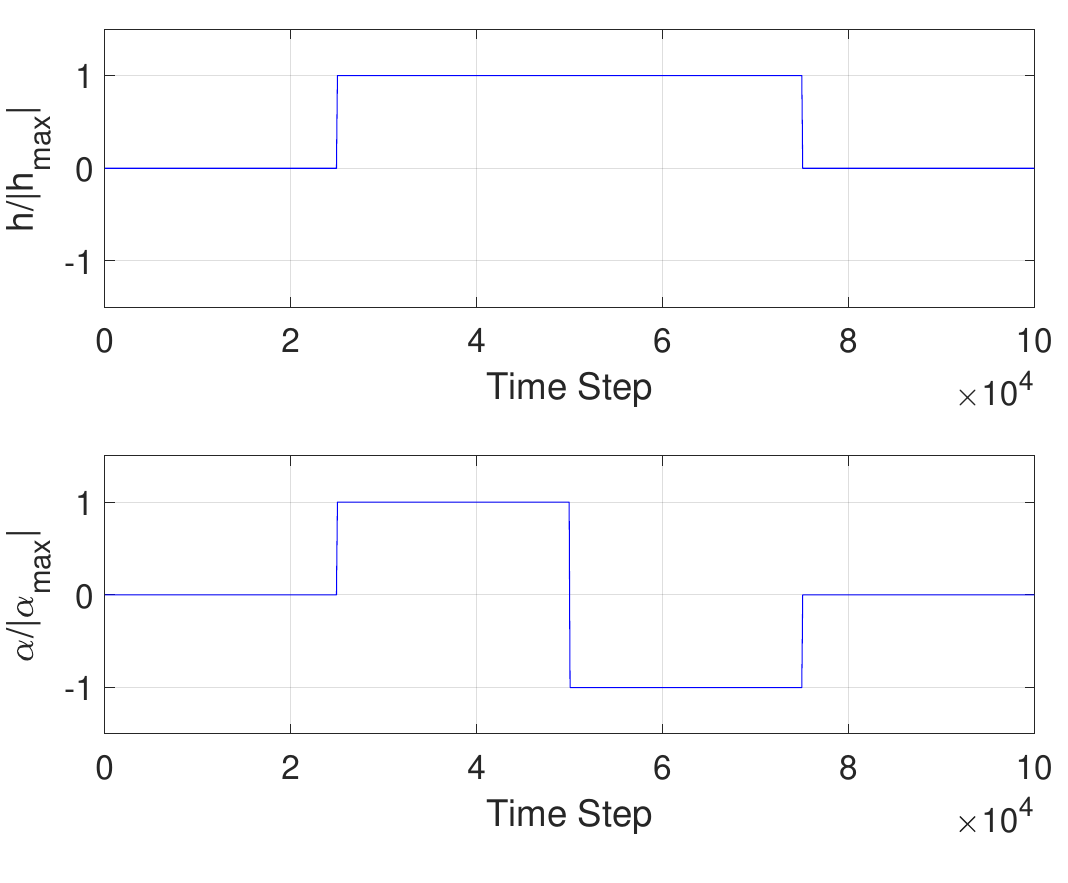}} 
    	 \caption{Normalized motions prescribed by different orthogonal inputs.}
    	 \label{fig:figure2}
    \end{center}
\end{figure}

The test case considered throughout this paper involves a NACA 0012 airfoil at $M_{\infty}=0.8$, \textit{i.e.}, in the transonic regime, and with a zero mean flow angle of attack, {\em i.e.}, $\alpha_{0}=0$\@. The structural parameters that define the problem are $a_h=-2.0$, $x_{\alpha}=1.8$, $r_\alpha=1.865$, $\mu=60$, $\omega_h=\omega_\alpha=100$ rad/s, and $\omega_r=\omega_\alpha$ is used as a reference. A steady-state solution at the mean flow condition is used to initialize the unsteady CFD run. Such solution is obtained by the same CFD code running in steady Euler mode with a variable time step method. Further discussion on this initial steady state solution can be found in Refs.\ \cite{marques2008numerical,azevedo2012efficient}. 

From the converged stationary solution, the mesh is rigidly moved in a prescribed pattern and a total of 100,000 time steps of unsteady flow are computed with a constant $\Delta \Bar{t}$ of 0.003 dimensionless time units. For the present work, discrete step inputs and Walsh functions are considered in order to prescribe the airfoil movement. It is important to emphasize that the step inputs are used in a mode-by-mode fashion, whereas the Walsh functions are used to simultaneously excite all the system modes. Furthermore, the maximum amplitudes considered here are 0.000001$c$ for the plunging mode and 0.0001 deg. for the pitching degree of freedom. The reason for choosing such low amplitudes, as discussed earlier, is to remain within the ``quasi-linear'' region around the nonlinear steady solution, and to allow the CFD code to accurately propagate the disturbances from the discrete motion. A detailed discussion of the limits of such modal displacements is also presented in Ref.\ \cite{marques2008numerical}.

Moreover, the five cases considered for the Walsh function inputs are illustrated in Fig.\ \ref{fig:figure2}. The zero-motion regions in the input signals attempt to guarantee that the CFD computation starts and ends with a solution that does not have any perturbations from the prescribed motion. The choice of these sets is based on the work of Ref.\ \cite{azevedo2013effects}. It is important to emphasize that the simulations with any set of Walsh functions require a single unsteady CFD run, whereas those with a discrete step (mode-by-mode approach) would require two simulations in the present case, \textit{i.e.}, one for each modal movement. The solutions provided by the CFD code are, then, used by a system identification routine developed in MATLAB$^{\copyright}$. In order to establish a comparison, both mode-by-mode and simultaneous simulations are analyzed with the power spectral density method. The aerodynamic transfer functions obtained in this way are compared to those presented in Refs.\ \cite{marques2008numerical,azevedo2013effects}. Lastly, the discrete values referring to each transfer function are interpolated using the first and second forms of the Eversman and Tewari polynomials in order to provide a suitable input for flutter analysis in the frequency domain.

\begin{table}[hbt!]
    \caption{System identification variables.} 
    \label{tab:table1}
	\begin{center} {
		\begin{tabular}{ccccc}
		\hline \hline
		Case & Inputs & Blocks & Sample & Overlap \\ \hline
        00   & Step  & 1      & 1      & 0       \\
        01   & WF1   & 2      & 2      & 0       \\
        02   & WF1   & 2      & 1      & 0       \\
        03   & WF2   & 3      & 3      & 0       \\
        04   & WF2   & 3      & 2      & 1       \\
        05   & WF2   & 3      & 1      & 0       \\
        06   & WF3   & 3      & 3      & 0       \\
        07   & WF3   & 3      & 2      & 1       \\
        08   & WF3   & 3      & 1      & 0       \\
        09   & WF4   & 4      & 4      & 0       \\
        10   & WF4   & 4      & 3      & 2       \\
        11   & WF4   & 4      & 2      & 1       \\
        12   & WF4   & 4      & 2      & 0       \\
        13   & WF4   & 4      & 1      & 0       \\
        14   & WF5   & 4      & 4      & 0       \\
        15   & WF5   & 4      & 3      & 2       \\
        16   & WF5   & 4      & 2      & 1       \\
        17   & WF5   & 4      & 2      & 0       \\
        18   & WF5   & 4      & 1      & 0       \\ \hline \hline
	    \end{tabular}}
	\end{center}
 \end{table}

\subsection{Aerodynamic Transfer Functions}

The validation of the implemented procedure, that uses power spectral densities, was achieved by comparing the transfer functions obtained with a single unsteady CFD run with the ones previously calculated using a mode-by-mode approach, as exposed in Ref.\ \cite{azevedo2012efficient}. Besides, when transforming the discrete aerodynamic time histories into frequency domain transfer functions, one is faced with certain variables regarding the size of the sampled data and the number of overlapping points. To simplify the present analysis, the samples and overlaps were determined as integer values of information blocks. An information block is defined here by the length of the input discrete data where no variations occur. As a result, the five unsteady CFD results were, then, spread through a total of 18 different test cases. The number of information blocks, size of the sample, and the number of overlapping blocks adopted in each test case, as well as the input it refers to, are described in Table \ref{tab:table1}. It also includes a test case referred to as Case 00 that applies discrete step inputs in a mode-by-mode fashion. Initially, the present work applies a rectangular window with the same size of the sampled data, but investigations conducted herein also analyze the effects of employing the Hanning window based on signal processing techniques. 
 
\begin{figure}[htb!]
    \begin{center}
    	 \subfigure[$G_{C_{l},h}$ results for Case 08 using rectangular window overlaid by Case 00 counterpart.]{ \includegraphics[scale=0.4,trim = 3.5cm 8.5cm 4.5cm 9cm,clip]{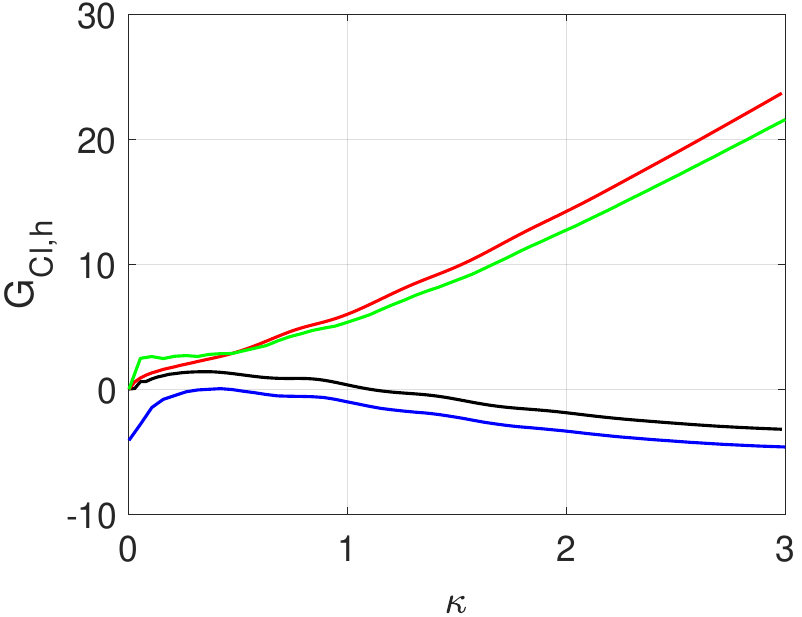}} \quad
    	 \subfigure[$G_{C_{l},\alpha}$ results for Case 08 using rectangular window overlaid by Case 00 counterpart.]{ \includegraphics[scale=0.4,trim = 3.5cm 8.5cm 4.5cm 9cm,clip]{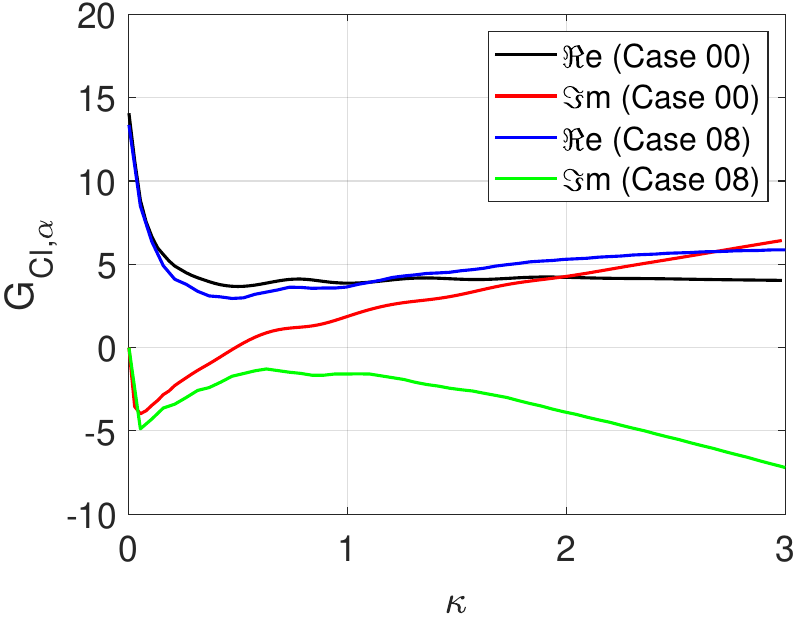}} \\
    	 \subfigure[$G_{C_{m},h}$ results for Case 08 using rectangular window overlaid by Case 00 counterpart.]{ \includegraphics[scale=0.4,trim = 3.5cm 8.5cm 4.5cm 9cm,clip]{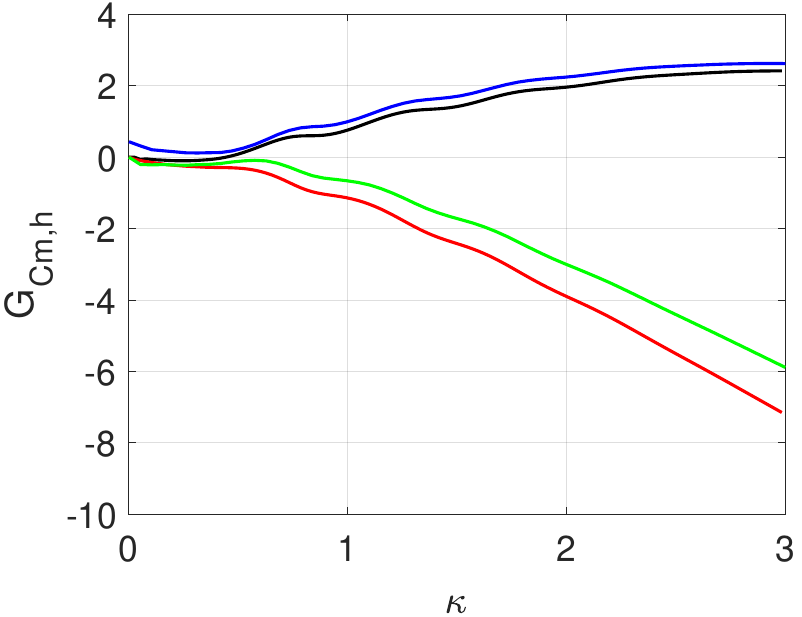}} \quad
    	\subfigure[$G_{C_{m},\alpha}$ results for Case 08 using rectangular window overlaid by Case 00 counterpart.]{ \includegraphics[scale=0.4,trim = 3.5cm 8.5cm 4.5cm 9cm,clip]{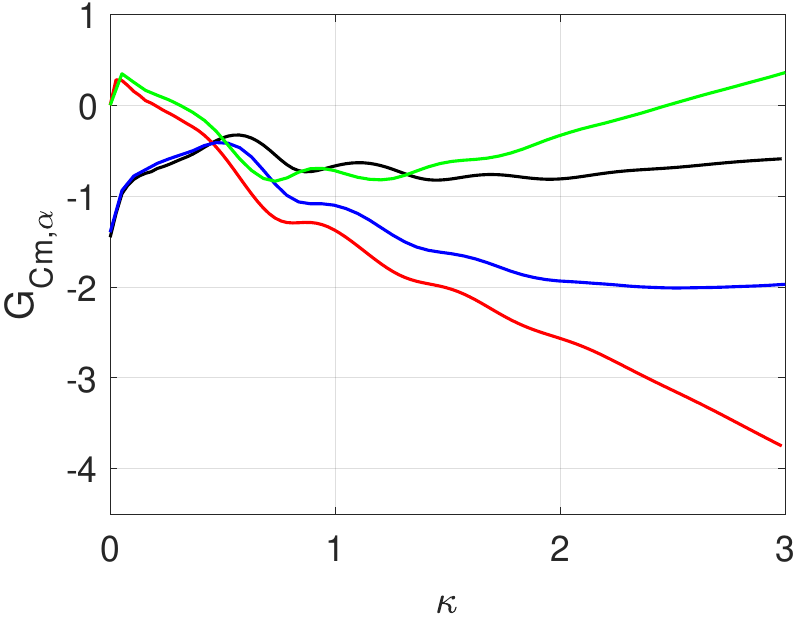}} 
    	 \caption{Aerodynamic transfer function for Case 08.}
    	 \label{fig:figure3}
    \end{center}
\end{figure}

Transfer functions are obtained for all the 18 test cases, but just some of the results are presented in this document for the sake of brevity. Initial tests indicate that the prescribed input signals in plunge and pitch modes and, also, their derivatives must be orthogonal to one another in order to apply the simultaneous excitation approach consistently. Such conclusion is based on the assessment of cross-power spectral densities between the input signals. The inputs WF1 and WF3 do not satisfy this linear independence relation either between the input signals or their derivatives. Consequently, results from these input signals are not capable of reproducing the transfer functions of the benchmark Case 00. One can observe this behavior in Fig.\ \ref{fig:figure3} that depicts the resulting transfer functions for Case 08, which is associated with WF3 input, when considering the rectangular window in the signal processing.
 
\begin{figure}[htb!]
    \begin{center}
    	 \subfigure[$G_{C_{l},h}$ results for Case 12 using rectangular window overlaid by Case 00 counterpart.]{ \includegraphics[scale=0.4,trim = 3.5cm 8.5cm 4.5cm 9cm,clip]{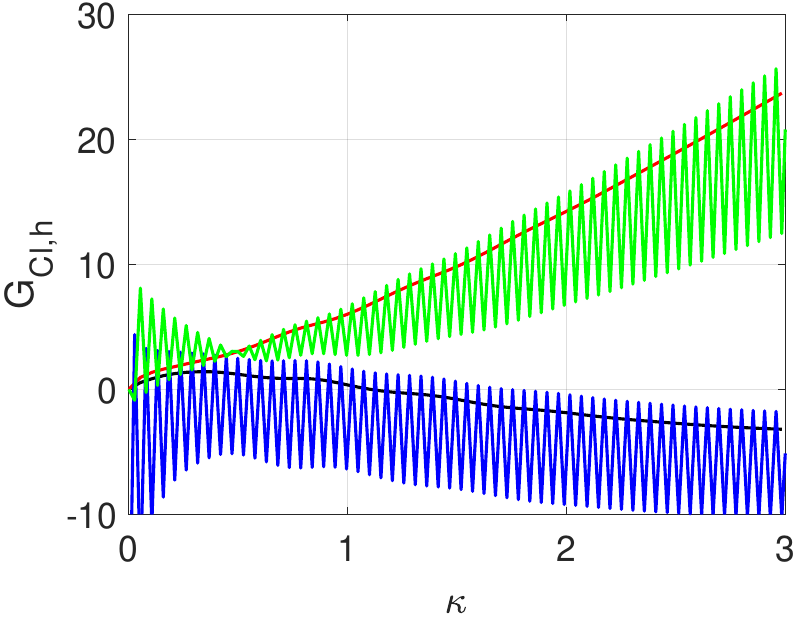}} \quad
    	 \subfigure[$G_{C_{l},\alpha}$ results for Case 12 using rectangular window overlaid by Case 00 counterpart.]{ \includegraphics[scale=0.4,trim = 3.5cm 8.5cm 4.5cm 9cm,clip]{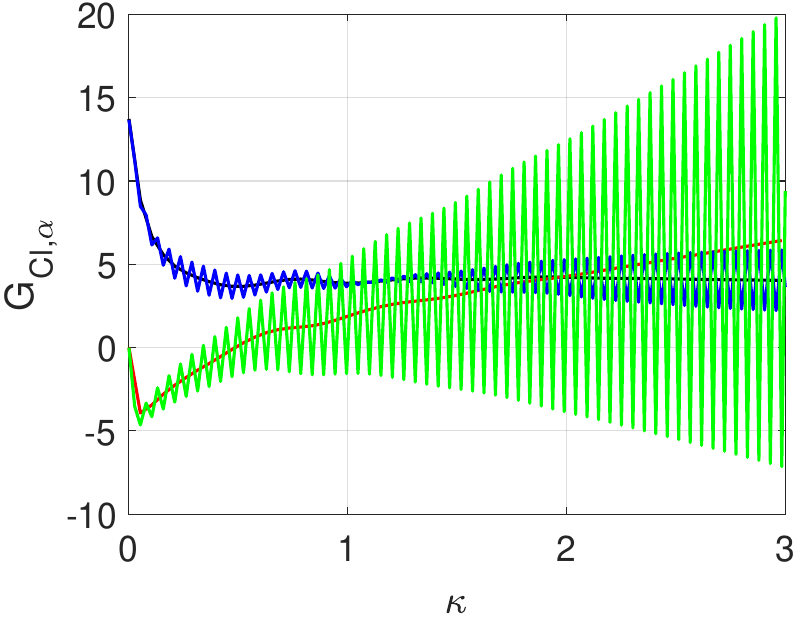}} \\
    	 \subfigure[$G_{C_{m},h}$ results for Case 12 using rectangular window overlaid by Case 00 counterpart.]{ \includegraphics[scale=0.4,trim = 3.5cm 8.5cm 4.5cm 9cm,clip]{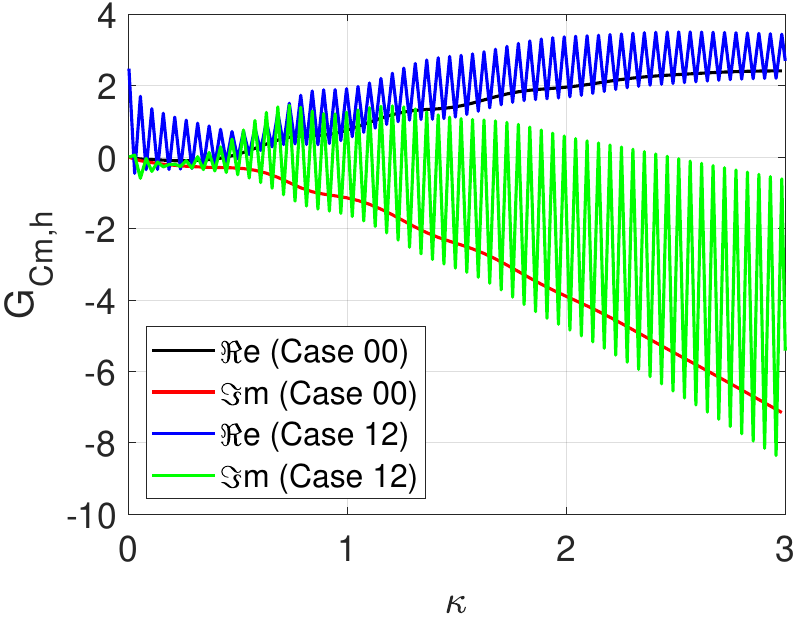}} \quad
    	\subfigure[$G_{C_{m},\alpha}$ results for Case 12 using rectangular window overlaid by Case 00 counterpart.]{ \includegraphics[scale=0.4,trim = 3.5cm 8.5cm 4.5cm 9cm,clip]{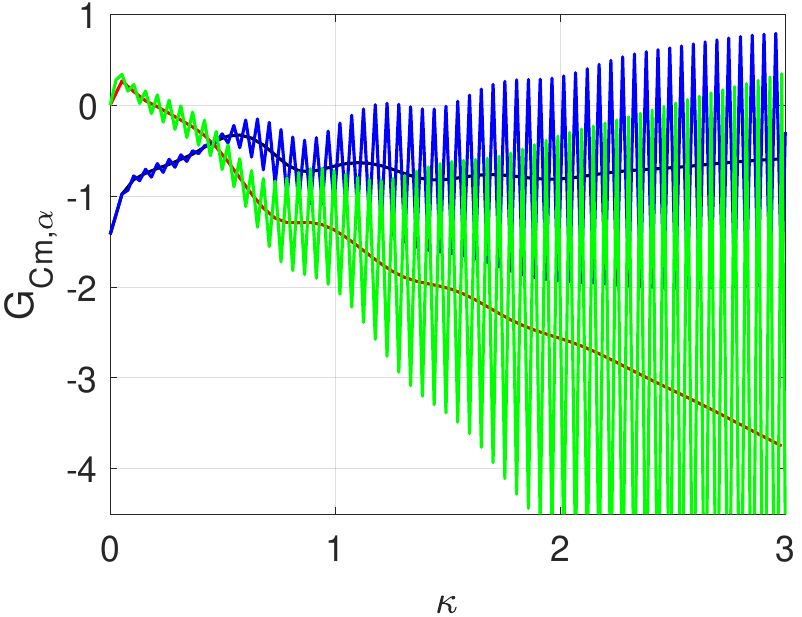}} 
    	 \caption{\label{fig:figure4}Aerodynamic transfer function for Case 12 using a rectangular window.}
    \end{center}
\end{figure}

\begin{figure}[htb!]
    \begin{center}
    	 \subfigure[$G_{C_{l},h}$ results for Case 12 using Hanning window overlaid by Case 00 counterpart.]{ \includegraphics[scale=0.4,trim = 3.5cm 8.5cm 4.5cm 9cm,clip]{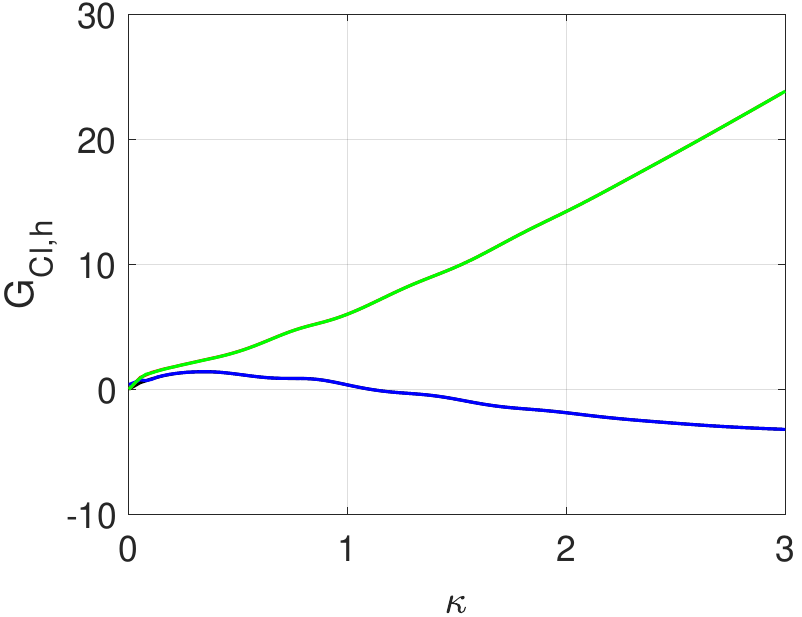}} \quad
    	 \subfigure[$G_{C_{l},\alpha}$ results for Case 12 using Hanning window overlaid by Case 00 counterpart.]{ \includegraphics[scale=0.4,trim = 3.5cm 8.5cm 4.5cm 9cm,clip]{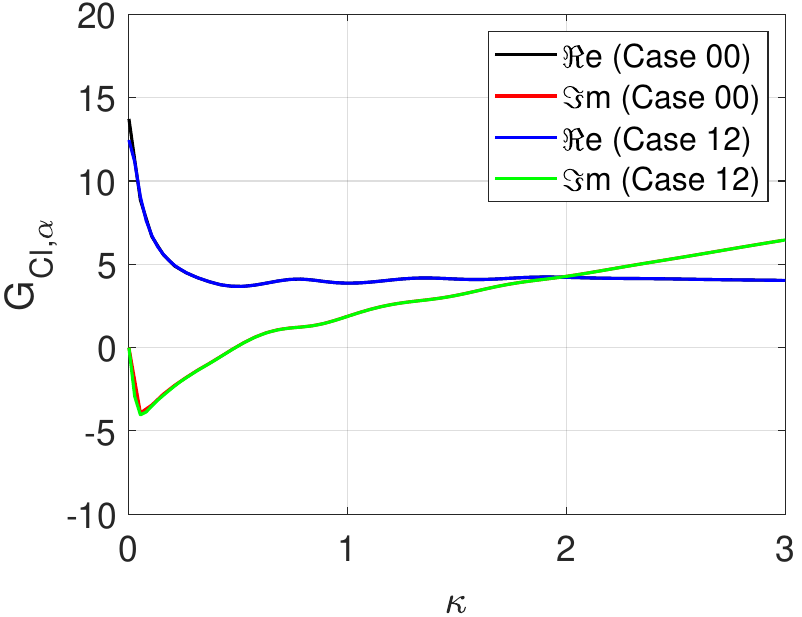}} \\
    	 \subfigure[$G_{C_{m},h}$ results for Case 12 using Hanning window overlaid by Case 00 counterpart.]{ \includegraphics[scale=0.4,trim = 3.5cm 8.5cm 4.5cm 9cm,clip]{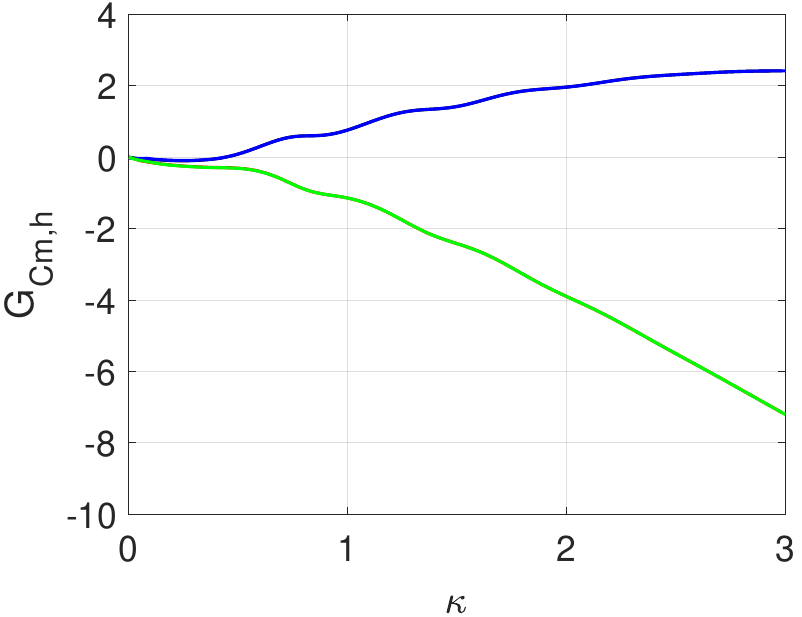}} \quad
    	\subfigure[$G_{C_{m},\alpha}$ results for Case 12 using Hanning window overlaid by Case 00 counterpart.]{ \includegraphics[scale=0.4,trim = 3.5cm 8.5cm 4.5cm 9cm,clip]{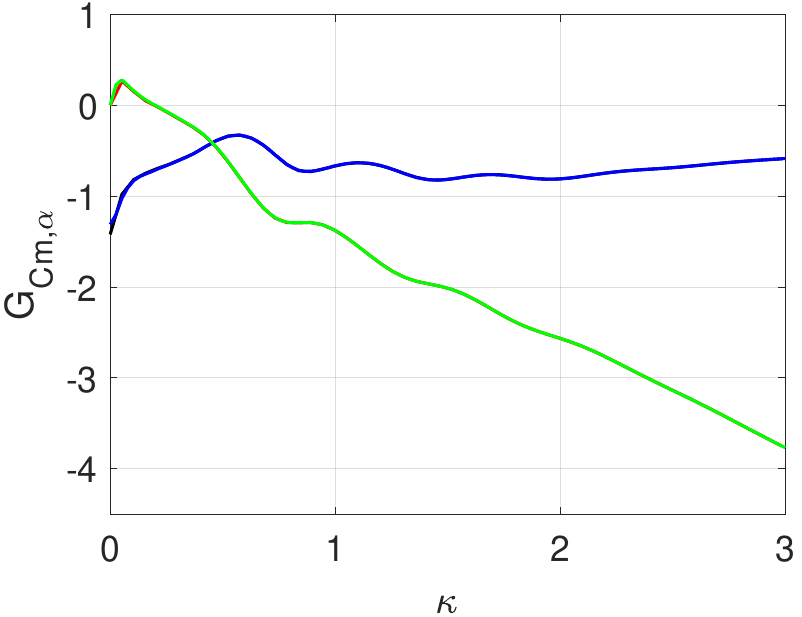}} 
    	 \caption{\label{fig:figure5}Aerodynamic transfer function for Case 12 using a Hanning window.}
    \end{center}
\end{figure}

The transfer functions for Case 12, when using rectangular window, are very oscillatory for both plunge and pitch modes, as illustrated in Fig.\ \ref{fig:figure4}. In general, this behavior is also present in results from other test sets, which are omitted here for brevity reasons. In view of signal processing techniques, this oscillatory trend is a consequence of the occurrence of the leakage phenomenon. More details about the leakage phenomenon can be found in Ref.\ \cite{oppenheim2001discrete}. On the other hand, Fig.\ \ref{fig:figure5} shows that there are no oscillations for both plunge and pitch modes when assessing Case 12 using a Hanning window. The application of the Hanning window in Case 12 solves the leakage problem because, by definition \cite{stoica2005spectral}, it nulls the contribution of the first and last data of the original signal. Hence, this result clearly indicates that the usefulness of different test cases in the context of a power spectral density analysis also depends on proper signal processing. Given the fact that Case 12 accordingly predicts the transfer functions when applying the Hanning window, this configuration is able to advance through the proposed procedure, in contrast to its counterpart with rectangular window. From now on, Case 13 is considered further in the present discussion instead of Case 12 for two reasons. Firstly, Case 13 is also capable of accurately predicting the aerodynamic transfer functions, similar to Case 12 in Fig. \ref{fig:figure5}. Additionally, results in general suggest that test cases related to a unit sample can better estimate flutter onset points compared to numerical results from Ref. \cite{rausch1990euler}. For these reasons, the following sections continue the proposed analysis presenting results from Case 13.

\FloatBarrier

\subsection{Polynomial Interpolation}

As mentioned earlier, the final stage in preparing the data for an aeroelastic stability analysis in the frequency domain consists in approximating the aerodynamic transfer function coefficients by the selected interpolating polynomial. Here, all calculations consider the first and second forms of the Eversman and Tewari polynomials \cite{eversman1991consistent} with the addition of 5 poles. The data fitting is performed for the reduced frequency range of $0 \leq \kappa \leq 3$ because it represents the needed aerodynamic information for the stability analyses in question. Moreover, given the fact that some of the proposed cases failed to produce a coherent set of transfer functions, the authors restricted the cases under evaluation for this stage. Initially, it is fundamental to ensure that the optimized poles, and the optimized linear coefficients by extension, that constitutes the interpolating polynomial are independent of the initial set of poles used to start the optimization process. Table \ref{tab:table2} shows the optimized poles using the first and second forms of Eversman and Tewari polynomials, respectively, and their corresponding flutter onset point for Case 13 assuming two distinct initial set of poles. In the present calculations, it is assumed that the flutter instability occurs at the smallest dynamic pressure at which the real part of one of the eigenvalues of the dynamic matrix becomes positive. More details regarding the flutter onset point identification are further discussed in the following subsection.

\begin{table}[b!]
    \caption{Optimized lag parameters of Case 13 with 1st and 2nd forms of the Eversman and Tewari polynomials.}
    \label{tab:table2}
    \small
    \begin{center} {
		\begin{tabular}{l|lc|lc} \hline \hline
         & \multicolumn{2}{c|}{First Form of Polynomials} & \multicolumn{2}{c}{Second Form of Polynomials} \\ \hline
        \multicolumn{1}{l|}{Initial poles, $\beta_n$} & Optimized poles, $\beta_n$ & U$^{*}_{f}$ & Optimized poles, $\beta_n$ & U$^{*}_{f}$ \\ \hline
        \multicolumn{1}{l|}{\{0.0, 0.375, 0.750, 1.125, 1.500\}}   & \{0.041, 0.416, 0.419, 0.416, 2.002\} & 4.73 & \{0.040, 0.559, 1.038, 0.579, 0.589\} & 4.79 \\
        \multicolumn{1}{l|}{\{1.5, 0.800, 0.350, 0.010, 2.100\}} & \{0.041, 0.419, 0.418, 0.415, 2.004\} & 4.73 & \{2.928, 0.619, 0.307, 0.031, 0.317\} & 4.55 \\ \hline \hline
	    \end{tabular}}
    \end{center}
\end{table}

\begin{figure}[htb!]
    \begin{center}
    	 \subfigure[$G_{C_{l},h}$ results for Case 13 layered over original data set.]{ \includegraphics[scale=0.4,trim = 3.5cm 8.5cm 4.5cm 9cm,clip]{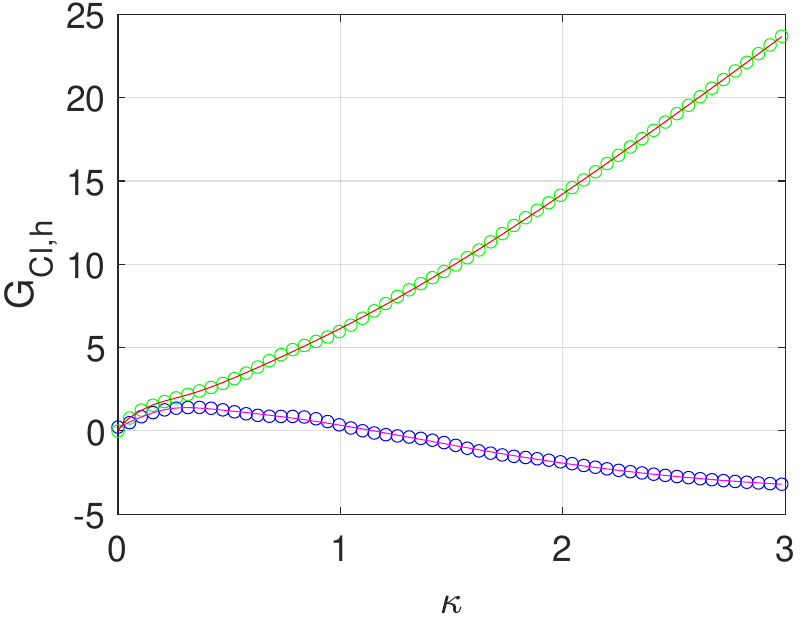}} \quad
    	 \subfigure[$G_{C_{l},\alpha}$ results for Case 13 layered over original data set.]{ \includegraphics[scale=0.4,trim = 3.5cm 8.5cm 4.5cm 9cm,clip]{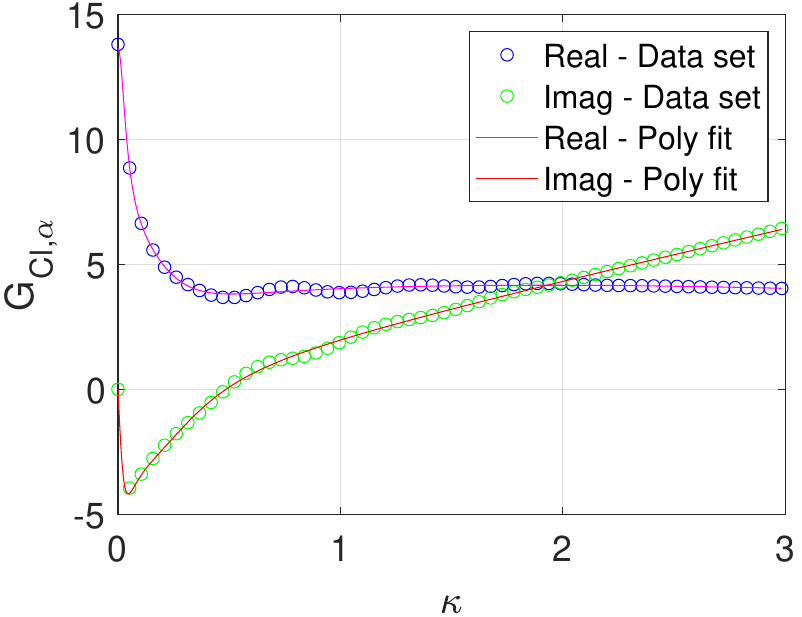}} \\
    	 \subfigure[$G_{C_{m},h}$ results for Case 13 layered over original data set.]{ \includegraphics[scale=0.4,trim = 3.5cm 8.5cm 4.5cm 9cm,clip]{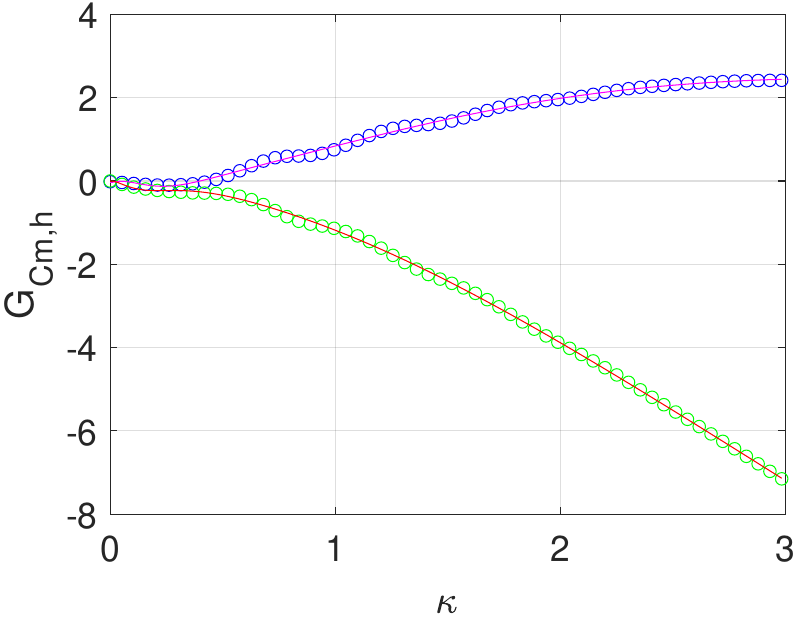}} \quad
    	\subfigure[$G_{C_{m},\alpha}$ results for Case 13 layered over original data set.]{ \includegraphics[scale=0.4,trim = 3.5cm 8.5cm 4.5cm 9cm,clip]{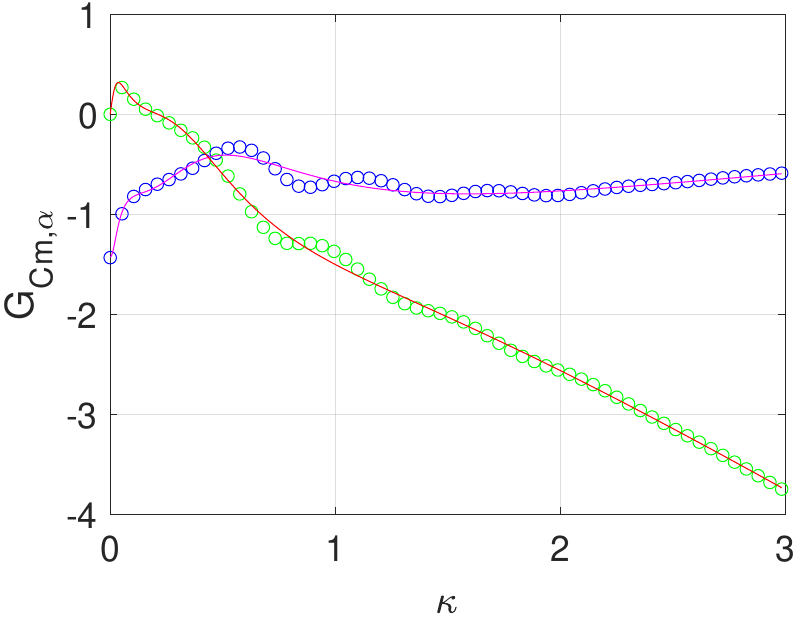}} 
    	 \caption{Comparison of aerodynamic transfer function behavior when the 2nd form of interpolating polynomials is used for Case 13.}
    	 \label{fig:figure6}
    \end{center}
\end{figure}

In spite of distinct initial poles, it is clear that the optimized poles are very similar and, hence, result in the same flutter onset point when applying the first form of the interpolating polynomials in Case 13, as indicated in Table \ref{tab:table2}. However, it is important to point out that this behavior is not observed for other test cases, such as Case 12, in which the flutter onset point varies from having a characteristic speed of $4.67$ to $4.61$ for the same initial set of poles as indicated in Table \ref{tab:table2}. The results in Table \ref{tab:table2} show a different behavior for Case 13, when the second form of the interpolating polynomials is used. In this case, the optimized poles of Case 13 also depend on the selected initial poles. The Nelder-Mead optimization algorithm converges to a function local minimum; however, it does not guarantee the identification of a global minimum. In addition, given the fact that the optimized poles are usually prone to depend on the initial poles, it is evident that the cost function of the optimization process is ill-posed. In other words, the aforementioned cost function demonstrates to have several local minima. To verify the fitting results, Fig.\ \ref{fig:figure6} presents a comparison of the transfer functions obtained for Case 13 when the 2nd form of the interpolating polynomials is used. The figure compares the RFA results with the original transfer function obtained directly from the CFD data. As one can see in Fig.\ \ref{fig:figure6}, in general, there is a good agreement between the sets of results from the approximating polynomials and the original transfer functions obtained directly from the CFD data. Discrepancies mainly occur in regions in which there is a rapid variation in the CFD results, which the interpolated RFAs do not seem to be able to completely follow. 

\begin{figure}[htb!]
    \begin{center}
    	 {\includegraphics[scale=0.47,trim = 3.5cm 8cm 4.5cm 9cm,clip]{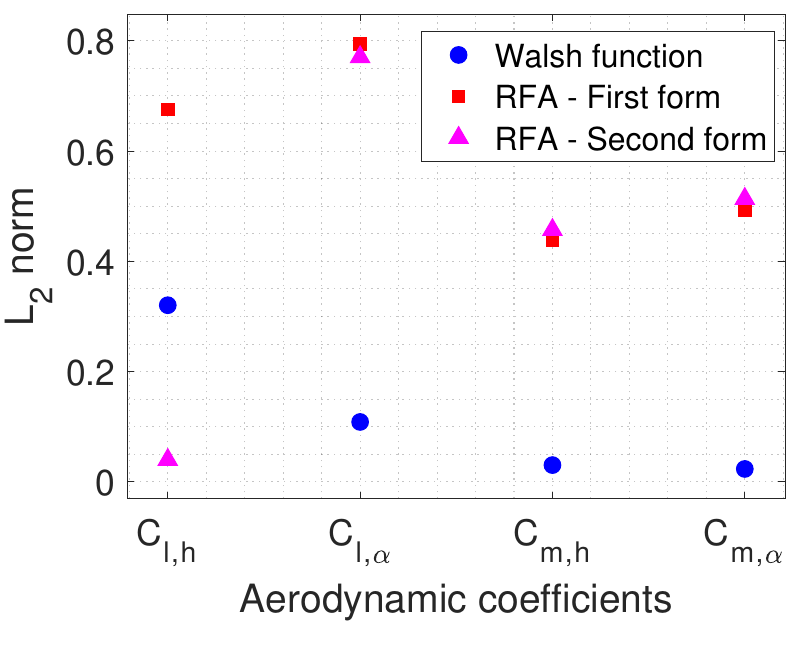}}
    	 \caption{L$_2$ norm of the difference between the aerodynamic transfer functions for Cases 00 and 13.}
    	 \label{fig:figure7}
    \end{center}
\end{figure}

It should be noticed that the flutter stability analysis is based on data from the approximated aerodynamic transfer functions. Therefore, it is appropriate to quantify the error associated with this approximation using the transfer functions obtained with the mode-by-mode approach as a reference. For this purpose, the present work has calculated the L$_2$ norms of the differences between the aerodynamic transfer functions for Cases 00 and 13\@. Three different comparisons are made. The first one simply compares the transfer functions, obtained directly from the CFD data, between Cases 00 and 13\@. In other words, in this comparison there is no approximation of the transfer functions by the interpolating polynomials. Therefore, this comparison is essentially measuring the effects of identifying the transfer functions with a simultaneous excitation as opposed to the mode-by-mode approach. The results for this case are labeled as ``Walsh function'' in Fig.\ \ref{fig:figure7}. The other two comparisons are directed towards evaluating the effects of the polynomial approximation process, {\em i.e.}, the creation of the RFAs. Hence, for these cases, the comparisons are between the transfer functions of Case 00 obtained directly from the CFD data and the RFAs for Case 13 obtained using the 1st and 2nd forms of the interpolating polynomials. These two cases are labeled as ``RFA - First form'' and ``RFA - Second form'' in Fig.\ \ref{fig:figure7}. As indicated, in all cases, the comparison consists in computing the L$_2$ norms of the differences between the transfer functions computed in the different ways.

Figure \ref{fig:figure7} indicates that the multiple input multiple output (MIMO) approach, associated with system identification techniques, does not introduce as much error in the aerodynamic transfer functions as the rational function approximations. For the particular test case shown here, an exception is the $C_{l,h}$ transfer function, since the second form of the interpolating polynomial yields a smaller error than the MIMO approach compared to Case 00. However, this is a fortuitous outcome, since other similar analyses using the test cases in Table \ref{tab:table1} have indicated that all RFA cases ended up having a larger error. Such results are indicating that, in general, the use of the MIMO approach or the mode-by-mode identification are essentially equivalent, whereas the error is actually introduced by the process of approximating the identified transfer functions with the rational polynomial functions. In other words, the rational function approximation is primarily responsible for introducing error in the flutter stability analysis. As the original authors have indicated \cite{eversman1991consistent}, the second form of the interpolating polynomial does not improve the correlation with the CFD data. However, it produces a much better-conditioned eigenvalue problem. Moreover, extrapolating the second form of the Eversman and Tewari polynomials using $n_\beta$ identical poles to describe the approximation has not improved the flutter onset point estimates for the test cases addressed in the present work.

\subsection{Aeroelastic Stability Analysis}

\begin{figure}[htb!]
    \begin{center}
    	 \subfigure[First structural mode.]{ \includegraphics[scale=0.47,trim = 3.5cm 8cm 4.5cm 9cm,clip]{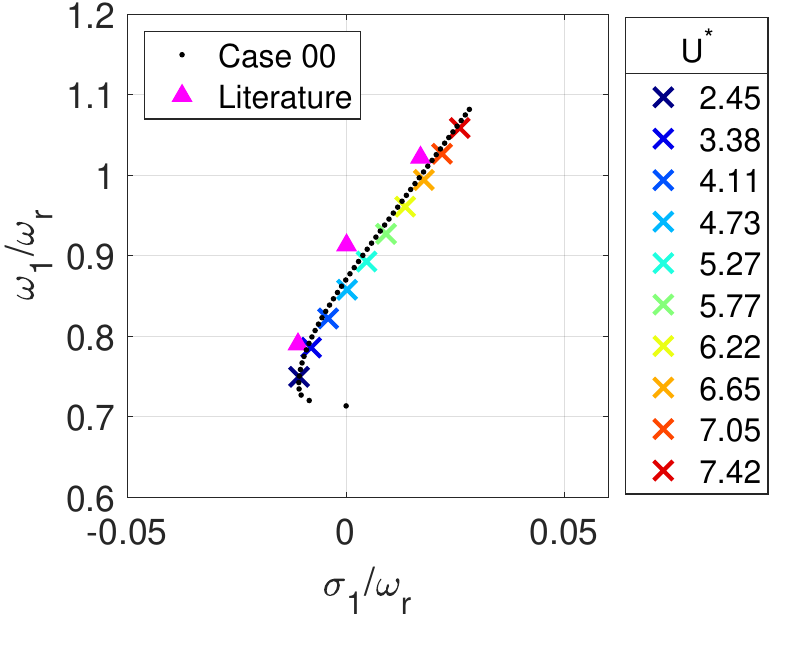}} \quad
    	 \subfigure[Second structural mode.]{ \includegraphics[scale=0.47,trim = 3.5cm 8cm 4.5cm 9cm,clip]{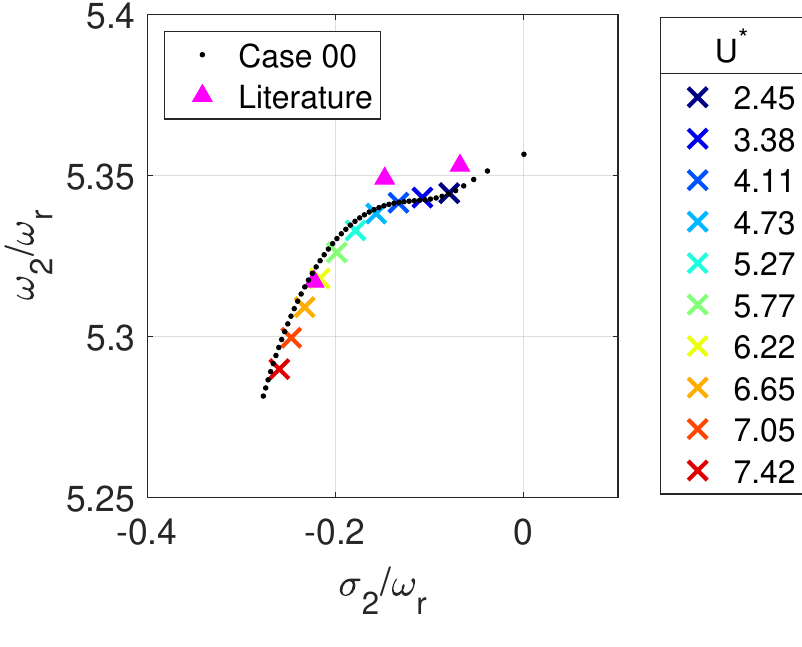}} 
    	 \caption{Root locus of the first and second structural modes for Case 13 using the first form of Eversman and Tewari interpolating polynomials with 5 poles.}
    	 \label{fig:figure8}
    \end{center}
\end{figure}

\begin{figure}[htb!]
    \begin{center}
    	 \subfigure[First structural mode.]{ \includegraphics[scale=0.47,trim = 3.5cm 8cm 4.5cm 9cm,clip]{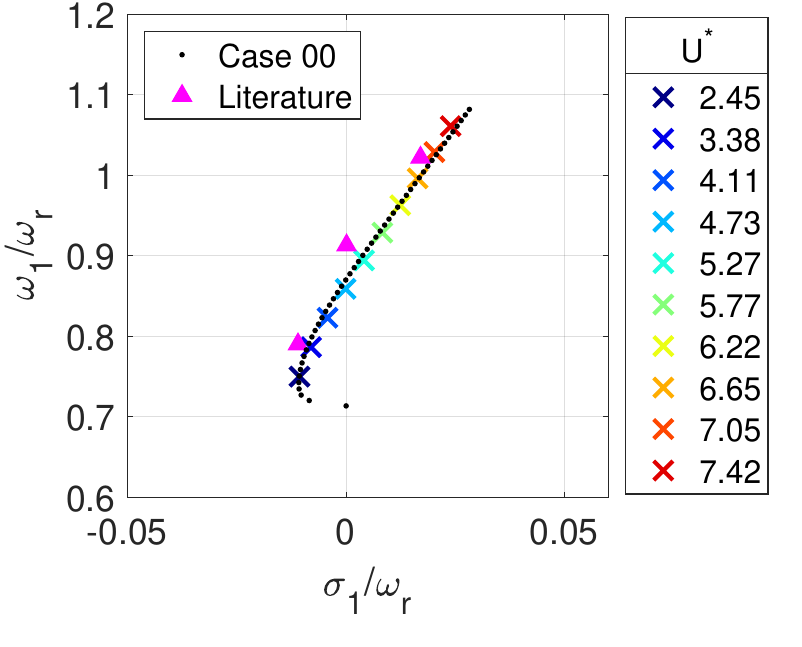}} \quad
    	 \subfigure[Second structural mode.]{ \includegraphics[scale=0.47,trim = 3.5cm 8cm 4.5cm 9cm,clip]{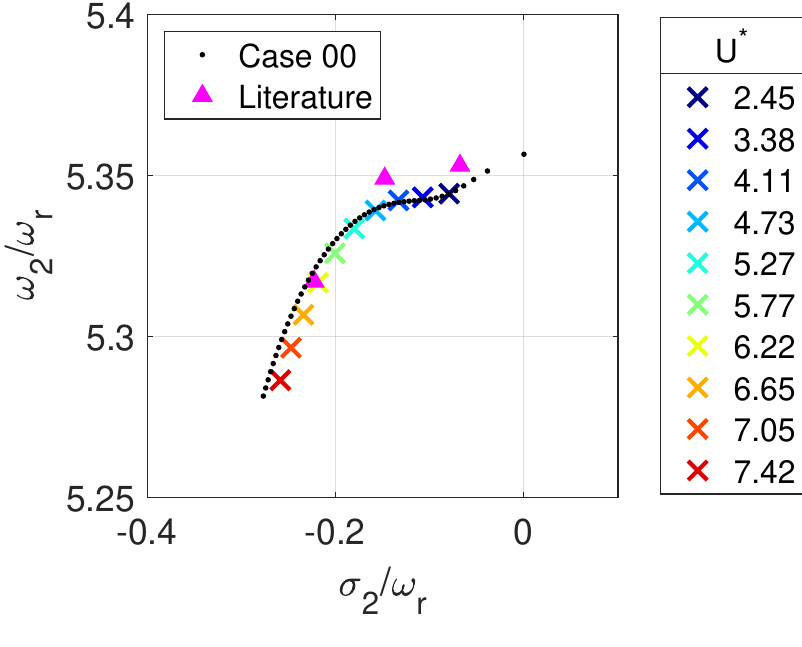}} 
    	 \caption{Root locus of the first and second structural modes for Case 13 using the second form of Eversman and Tewari interpolating polynomials with 5 poles.}
    	 \label{fig:figure9}
    \end{center}
\end{figure}

Finally, the obtained polynomials are employed in the solution of the flutter stability eigenvalue problem. The stability root loci of both structural modes, associated with the first and second forms of the interpolating polynomials in Case 13, can be seen in Figs.\ \ref{fig:figure8} and \ref{fig:figure9}, respectively. These figures also present numerical results from Case 00 and those from Ref.\ \cite{rausch1990euler}. In the search for the flutter onset point, the characteristic dynamic pressure parameter is varied from $Q^{*}=0.0$ up to 1.0, in $\Delta Q^{*}=0.01$ intervals. Each plotted point of characteristic speed corresponds to one of these values. The literature data, however, are only available for $Q^{*}=$ 0.2, 0.5, and 0.8\@. Obviously, not all the calculated eigenvalues are indicated in the figures in order to avoid cluttering them. However, the values shown for the present calculations allow for a visual indication of the flutter condition.

Figures \ref{fig:figure8} and \ref{fig:figure9} indicate that the modal damping and frequency behavior of the aeroelastic modes predicted by the proposed procedure using the mode-by-mode approach and the Walsh functions are very similar. The behavior of the aeroelastic modes seems to also follow essentially the same behavior indicated in the results in Ref.\ \cite{rausch1990euler}. The differences seem to be consistent with the observed differences in the flutter dynamics pressure, or characteristic speed, for this case. An important aspect to be considered in the root locus analysis is that one is mostly concerned with the flutter onset point. It means that the characteristic speed associated with the flutter phenomenon is as relevant as the accuracy of all data in the root locus plot. Reference \cite{rausch1990euler} states that flutter occurs close to $Q^{*}=0.5$, which is a near neutrally stable condition for the same aeroelastic configuration considered herein. Such condition corresponds to a reference characteristic speed of $U^{*}=5.4772$. One can further observe from these figures that, as expected, there is not much difference between the results obtained with the 1st or the 2nd form of the interpolating polynomials.

Considering the literature data in Ref.\ \cite{rausch1990euler} as benchmark, Tables \ref{tab:table3} and \ref{tab:table4} present the modal damping and frequency values at the flutter onset point for the test cases from Table \ref{tab:table1} which are capable of reproducing the transfer functions with 5 poles, and with either rectangular or Hanning windows. By definition, the flutter onset point can be identified as the point at which the damping of one of the aeroelastic modes becomes identically zero. Results displayed in Tables \ref{tab:table3} and \ref{tab:table4}, however, present the lowest characteristic speed in which one of the dynamic matrix eigenvalues has a positive real part. Therefore, such condition is not exactly the flutter point, that is, when the damping is identically zero, but it is close enough for the present purposes, since the increment in the search process in the dynamic pressure parameter is very small. In other words, given the fact that $\Delta Q^{*}=0.01$, the identified results for the flutter onset point are considered to be sufficiently accurate. The percentage error indicated in Tables \ref{tab:table3} and \ref{tab:table4} compares the flutter characteristic speed, U$^{*}_{f}$, with the value obtained for this flight condition in Ref.\ \cite{rausch1990euler}. All flutter onset point results calculated in the present work are conservative. This means that the present results yield flutter characteristic speeds which are lower than the reference value.

Moreover, the test cases in which the samples used in the system identification process consisted of a single data block, as defined with regard to Table \ref{tab:table1}, typically yield lower percentage errors for each set of Walsh function inputs. For instance, if we concentrate on the results with the 2nd form of the interpolating polynomials, one can see that cases 13 and 18, as listed in Table \ref{tab:table4}, have lower errors with respect to the literature data than the other tests cases using the WF4 and WF5 sets of Walsh functions, respectively. Moreover, typically again, those test cases do not require the use of Hanning windows in the system identification process. In other words, one could use simple rectangular windows in the identification process.

\begin{table}[htb!]
\caption{\label{tab:table3} Flutter onset points for NACA 0012 airfoil at $M_{\infty}=0.80$ for different test cases with first form of Eversman and Tewari interpolating polynomials and $n_\beta=5$ poles.}
    \begin{center} {
		\begin{tabular}{cccccccc}
		\hline
        Input & Case & $\sigma_1/\omega_r$ & $\omega_1/\omega_r$ & $\sigma_2/\omega_r$ & $\omega_2/\omega_r$ & U$^{*}_{f}$ & Error \%  \\ \hline
        Step  & 00  & 3.57$\times$10$^{-6}$ & 0.87  & -0.17 & 5.34  & 4.98 & 9.0 \\
        WF2 & 04 & 4.14$\times 10^{-4}$ & 0.86 & -0.16 & 5.34 & 4.79 & 12.6 \\
        WF2 & 05 & 1.63$\times 10^{-4}$ & 0.86 & -0.16 & 5.34 & 4.79 & 12.6 \\
        WF4 & 11 & 3.58$\times 10^{-4}$ & 0.86 & -0.16 & 5.34 & 4.73 & 13.7 \\
        WF4 & 12 & 4.21$\times 10^{-4}$ & 0.85 & -0.16 & 5.34 & 4.67 & 14.7 \\
        WF4 & 13 & 2.17$\times 10^{-4}$ & 0.86 & -0.16 & 5.34 & 4.73 & 13.7 \\
        WF5 & 16 & 2.70$\times 10^{-4}$ & 0.82 & -0.13 & 5.34 & 4.05 & 26.1 \\
        WF5 & 17 & 1.19$\times 10^{-4}$ & 0.82 & -0.13 & 5.34 & 4.05 & 26.1 \\
        WF5 & 18 & 3.03$\times 10^{-4}$ & 0.88 & -0.17 & 5.34 & 4.95 & 9.5  \\
        \hline
	    \end{tabular}}
    \end{center}
\end{table}

\begin{table}[htb!]
\caption{\label{tab:table4} Flutter points for NACA 0012 airfoil at $M_{\infty}=0.80$ for different test cases with second form of Eversman and Tewari interpolating polynomials and $n_\beta=5$ poles.} 
    \begin{center} {
		\begin{tabular}{cccccccc}
		\hline
        Input & Case & $\sigma_1/\omega_r$ & $\omega_1/\omega_r$ & $\sigma_2/\omega_r$ & $\omega_2/\omega_r$ & U$^{*}_{f}$ & Error \%  \\ \hline
        Step  & 00  & 3.57$\times$10$^{-6}$ & 0.87  & -0.17 & 5.34  & 4.98 & 9.0 \\
        WF2 & 04 & 3.97$\times 10^{-4}$ & 0.86 & -0.16 & 5.34 & 4.79 & 12.6 \\
        WF2 & 05 & 1.83$\times 10^{-4}$ & 0.86 & -0.16 & 5.34 & 4.73 & 13.7 \\
        WF4 & 11 & 4.14$\times 10^{-4}$ & 0.86 & -0.16 & 5.34 & 4.73 & 13.7 \\
        WF4 & 12 & 5.19$\times 10^{-4}$ & 0.84 & -0.15 & 5.34 & 4.49 & 18.0 \\
        WF4 & 13 & 2.98$\times 10^{-4}$ & 0.86 & -0.16 & 5.34 & 4.79 & 12.6 \\
        WF5 & 16 & 2.99$\times 10^{-4}$ & 0.82 & -0.13 & 5.34 & 4.05 & 26.1 \\
        WF5 & 17 & 1.25$\times 10^{-4}$ & 0.82 & -0.13 & 5.34 & 4.05 & 26.1 \\
        WF5 & 18 & 2.47$\times 10^{-4}$ & 0.87 & -0.17 & 5.34 & 5.01 & 8.5  \\
        \hline
	    \end{tabular}}
    \end{center}
    \vspace{-0.4cm}
\end{table}

\vspace{-0.1cm}
\section{Concluding Remarks}

The authors present a detailed analysis of the construction of reduced-order models capable of performing transonic aeroelastic stability analyses in the frequency domain with a unique unsteady CFD run. Results are presented for a NACA 0012 airfoil-based typical section with freestream conditions $M_{\infty}=0.8$ and $\alpha_{0}=0$ deg. It is initially shown that an accurate identification of the aerodynamic transfer functions depends on two aspects, namely the orthogonality of the input signals and their derivatives to one another, and proper signal processing. Results show that the investigated CFD-based reduced-order models are capable of consistently addressing transonic aeroelastic analysis in the context of small displacements in modal coordinates. 

However, it is not possible to ensure that the non-gradient optimization process, which essentially computes the aerodynamic lags and polynomial coefficients for the construction of the rational function approximations, can consistently converge to the same poles for a given flight condition regardless of the numerical control parameters used. Additionally, such interpolating polynomials, which attempt to represent the aerodynamic transfer functions in the continuous domain, are primarily responsible for introducing errors in the aeroelastic stability analyses. Such additional errors do not invalidate the proposed procedure, because of their low magnitude, but they can certainly affect the flutter onset point identification. 

The present effort highlights the reliability of the simultaneous excitation approach, which allows a unique unsteady CFD run and, consequently, significant computational cost savings for the creation of the aerodynamic reduced order model (ROM). Although the test case used in the present work is a very classic and simple one, the capability of generating the aerodynamic ROM for aeroelastic analysis with a single unsteady CFD run is particularly relevant when addressing complex engineering problems. It is also important to emphasize that, despite the difficulties discussed with regard to the optimization process, the procedure presented here can identify the flutter onset point with percentage errors as low as 8.5\% in comparison to the numerical benchmark case from literature. Another aspect is that, if one assumes the mode-by-mode approach as a reference, these percentage errors are as low as 0.6\%. Finally, all the results presented herein are conservative because the flutter speeds are all below the reference literature value.

\vspace{-0.3cm}
\section*{Acknowledgments}
The authors gratefully acknowledge the support for this research provided by Fundação de Amparo à Pesquisa do Estado de São Paulo, FAPESP, through a Master of Science Scholarship for the first author according to FAPESP Process No.\ 2022/01397-0\@. This work was also supported by Fundação Coordenação de Aperfeiçoamento de Pessoal de Nível Superior, CAPES, under the Research Grants No.\ 88887.634461/2021-00 and 88887.609895/2021-00\@. Additional support provided by FAPESP under the Research Grant No.\ 2013/07375-0 is also gratefully acknowledged. The present work has also received support from Conselho Nacional de Desenvolvimento Cient\'{\i}fico e Tecnol\'{o}gico, CNPq, under the Research Grant No.\ 309985/2013-7\@. This study was financed in part by the Coordena\c{c}\~{a}o de Aperfei\c{c}oamento de Pessoal de N\'{\i}vel Superior - Brasil (CAPES) - Finance Code 001\@.

\bibliographystyle{plain}
\bibliography{biblio}

\end{document}